\documentclass[11pt]{article}

\usepackage[comma,sort]{natbib}
\usepackage{url}

\usepackage[margin=1in]{geometry}

\usepackage{bm}
\usepackage{amsmath}
\usepackage{amsthm}
\usepackage{amsfonts}
\usepackage{amssymb}
\usepackage{shortcutsg}
\usepackage[capitalize]{cleveref}
\Crefname{assumption}{Assumption}{Assumptions}
\Crefname{appsec}{Appendix}{Appendices}

\usepackage{appendix}

\newtheorem{theorem}{Theorem}

\theoremstyle{definition}
\newtheorem{example}{Example}
\newtheorem{assumption}{Assumption}

\title{Demistifying Inference after Adaptive Experiments}
\author{Aur\'elien Bibaut and Nathan Kallus}
\date{}

\begin{document}

\maketitle

\begin{abstract}
Adaptive experiments such as multi-arm bandits adapt the treatment-allocation policy and/or the decision to stop the experiment to the data observed so far. This has the potential to improve outcomes for study participants within the experiment, to improve the chance of identifying best treatments after the experiment, and to avoid wasting data. Seen as an experiment (rather than just a continually optimizing system) it is still desirable to draw statistical inferences with frequentist guarantees. The concentration inequalities and union bounds that generally underlie adaptive experimentation algorithms can yield overly conservative inferences, but at the same time the asymptotic normality we would usually appeal to in non-adaptive settings can be imperiled by adaptivity. In this article we aim to explain why, how, and when adaptivity is in fact an issue for inference and, when it is, understand the various ways to fix it: reweighting to stabilize variances and recover asymptotic normality, always-valid inference based on joint normality of an asymptotic limiting sequence, and characterizing and inverting the non-normal distributions induced by adaptivity.
\end{abstract}

\section{Introduction}

Adapting sampling policies to past data is important whenever data is expensive to collect, whether in actual incurred costs (money or effort) or in opportunity cost (a sale we could have made or human life we could have improved or saves). Bandit and (online) reinforcement learning (RL) algorithms are commonplace in recommendation \citep[\eg,][]{li2010contextual} and revenue management \citep[\eg,][]{qiang2016dynamic,kallus2020dynamic}, where adapting decisions means we are continually improving our recommendations or pricing while these algorithms are running in production. Often in these applications we may not have a specific scientific question to answer and simply want algorithms that recommend or price well, \eg, for maximum cumulative revenue. Increasingly, however, bandits and RL are used in experiments where we do want to infer something from the data. Real-world examples of such experiments include \citet{cohen2015price,athey2022contextual,caria2023adaptive,offer2020optimal}.

Adaptation for the sake of experimentation is appealing both because it can allow for better outcomes for participants during the experiment and because it can improve the chance of identifying best treatments after the experiment. Besides, we might be interested in interpreting as an experiment any data collected by running a bandit even if originally for the sake of maximizing cumulative revenue. In all cases, in experiments, we are interested in answering a scientific question, and for this purpose we may be interested in statistical inference and in particular confidence intervals with frequentist guarantees for coverage and power. For example, such confidence intervals can offer credible evidence for whether a novel policy we learned has any improvement over a baseline.

When treatment-allocation policies and sampling stopping time are fixed in advance, inference is often based on inverting the central limit theorem (CLT), possibly with techniques to deal with unknown nuisances to be estimated \citep[\eg][]{chernozhukov2024applied}. However, when the treatment-allocation policy and/or sampling stopping time are adaptive, such approaches to inference can be imperiled, as highlighted by a variety of recent work. 

The purpose of this article is to explain why, how, and when adaptivity is in fact an issue for inference and, when it is, understand the various ways to fix it. There are practically relevant settings where adaptivity ``washes away" and becomes an insignificant artifact, so that even if sampling is on its face adaptive it is effectively non-adaptive and (martingale-)CLT-based inference is unchanged. When adaptivity remains significant through the experiment, we explain how the (nearly tight) conditions of self-normalized martingale CLTs fail. We then review three types of fixes. The first is based on reweighting to stabilize variances and recover the conditions of martingale CLT, enabling Wald-type confidence intervals based on asymptotic normality. The second is based on always-valid inference, which by being valid at all times is also valid at an adaptive stopping time. The third is based on characterizing and inverting the non-normal distributions induced by adaptive policies.

\section{Set-up}

\subsection{Data} We aim to analyze data sequences generated by the interaction of an environment and an experimenter/agent. The environment first produces an initial observation vector $L_0$. Over successive rounds $t=1,2,\ldots$ the agent first chooses an action $A_t$ from an action set $\mathcal{A}$, and the environment then produces an observation vector $L_t$. 
For any $t \geq 1$, let $\bar{O}_t = (L_0,A_1,L_1,\ldots, A_t, L_t)$. For any sequence $(l_0,a_1,l_1,\ldots)$, we denote by $a_t^-$ the finite subsequence of terms preceding $a_t$, that is $a_t^-= (l_0,\ldots, a_{t-1}, l_{t-1})$, and similarly $l_t^- = (l_0,\ldots,a_t)$. Denote by $P^{(t)}$ the distribution of $\bar O_t$.

\subsection{Statistical model} Let $\mu_\mathcal{A}$, $\mu_\mathcal{L}$ be measures over $\mathcal{A}$ and $\mathcal{L}$. We suppose that for every $t$, $P^{(t)}$ admits a density with respect to $\mu^t = \mu_{\Lcal}\times(\mu_{\Acal}\times\mu_{\Lcal})^t$. From the chain rule, any such density can be factorized as
\begin{align*}
    p^{(t)}(\bar o_t) = q_0(l_0) \prod_{s=1}^t g_s(a_s \mid a_s^-) q_s(l_s \mid l_s^-).
\end{align*}
For every $t \geq 1$, let $c_t$ be a mapping $\mathcal{L} \times (\mathcal{A} \times \mathcal{L})^{t-1} \to \mathcal{C}$ for some set $\Ccal$. For every $t \geq 1$, let $C_t = c_t(L_t^-)$. We will interpret $C_t$ as a summary of the history preceding action $L_t$. 
We restrict the class of densities we consider by assuming that (1) $L_t$ given the past only depends on the history summary $C_t$ and that (2) the conditional density $q_t$ of $L_t$ given $C_t$ is identical across all time points, that is that there exists a conditional density $q$ such that $q_t = q$ for all $t$. That is, we assume that the above density is of the form
\begin{align*}
    p^{(t)}(\bar o_t) = q_0(l_0) \prod_{t=1}^\infty g_s(a_s \mid a_s^-) q(l_s \mid c_s(l_t^-)).
\end{align*}

\begin{example}[Multi-armed bandit]\label{ex:mab} For the (stochastic) $K$-armed bandit model, we restrict $\Lcal=\RR$, $\Acal=[K]$, $L_0=0$, and
$q(l_t \mid l_t^-) = \widetilde q(l_t \mid a_t)$ for some collection of $K$ outcome densities $\widetilde q(\cdot \mid a)$, $a \in [K]$. That is, we equate $L_t$ with the reward at the end of round $t$. Note the design $g_t$ is unrestricted. We can further consider the many-arm setting (\eg, continuous arms) by simply removing the restriction on $\Acal$.\end{example}

\begin{example}[Contextual bandit]\label{ex:cb} For the (stochastic) contextual bandit model, we fix some context space $\Scal$, we set $\Lcal=\RR\times\Scal$ with base measure $\mu_\Lcal=\mu_Y\times\mu_\Scal$, we write $L_t=(Y_t,S_{t+1})$ and restrict $L_0=(0,S_1)$, and for $t\geq1$ we restrict $q(y_t,s_{t+1} \mid l_t^-) = \widetilde q(y_t \mid s_t,a_t)q_0(s_{t+1})$ for some arm-and-context-conditional density $\widetilde q(\cdot \mid s,a)$ (where we identify $q_0(s)=q_0(0,s)$). That is, $L_t$ consists of the reward at the end of round $t$ and the context at the beginning of round $t+1$.\end{example}

\begin{example}[Markov Decision Process]\label{ex:mdp} For a Markov decision process, we consider the same set up as the contextual bandit example except that for $t\geq1$ we restrict $q(l_t \mid l_t^-) = \widetilde q(y_t \mid a_t)\widetilde q_{\mathrm{transition}}(s_{t+1}\mid s_t,a_t)$ for some conditional density $\widetilde q_{\mathrm{transition}}$. That is, contexts are treated instead as states that evolve according to a time-homogeneous transition kernel.\end{example}

\begin{example}[Batched multi-armed bandit]\label{ex:batch} In the batched $K$-armed bandit setting, the design is updated over successive batches based on the realizations of the previous batches. For simplicity, consider the setting where each batch has size $n$. At the beginning of data collection for batch $1$, treatment assignments are drawn i.i.d. for units $i=1,\ldots,n$ following a fixed design $\widetilde g_1$: $A_{1,1},\ldots,A_{1,n} \sim \widetilde g_1$. Then, outcomes $L_{1,1},\ldots,L_{1,n}$ are drawn i.i.d. conditional on treatment assignments following outcome density $q$: for each $i=1,\ldots,n$, $L_{1,i} \mid A_{1,i} \sim q(\cdot \mid A_{1,i})$. Denote $A_b = (A_{b,1},\ldots, A_{b,n})$, $L_b = (L_{b,1},\ldots,L_{b,n})$, $O_b = (A_b, L_b)$. Treatment assignments $A_{b,1},\ldots, A_{b,n}$ in batch $b$ are drawn i.i.d. following a common density $\widetilde g_b$ that can depend on previous batch observations $O_1,\ldots, O_{b-1}$: $A_{b,1},\ldots, A_{b,n} \mid O_1,\ldots O_{b-1} \sim^{i.i.d.} \widetilde g_b(\cdot \mid O_1,\ldots O_{b-1})$. Outcomes in batch $b$ are drawn conditional on treatment assignments following conditional density $q$. Data collection stops at batch $B$ defined as a stopping time adapted to the filtration $(\mathcal{F}_b)_{b \geq 1}$ where $\mathcal{F}_b = \sigma(O_1,\ldots, O_b)$.\end{example}

\subsection{Target parameter and inference goal}
The statistical goal we consider in this paper is inference for a one-dimensional parameter $\theta(q)$ of the repeated factor $q$ of the density, at a potentially adaptive sample size time $\tau$, where we define $\tau$ as an $(\mathcal{F}_t)_{t \geq 0}$-adapted stopping time, where $\mathcal{F}_t = \sigma(\bar{O}_t)$. 

\begin{example}[Bandit arm means, treatment effects, and policy values]Consider the setting in \cref{ex:mab}. For a given vector $w\in\R K$, set $\theta(q)=\sum_{k=1}^Kw_k\int y\widetilde q(y\mid k)\,\mathrm{d}\mu_\Lcal(y)$. Usually, $w$ is a ``one-hot" vector ($\theta(q)$ is mean of an arm of interest) or the difference of two ``one-hot" vectors ($\theta(q)$ is an average treatment effect). For the setting in \cref{ex:cb}, we can consider 
\begin{align}\label{eq:policy value}\theta(q)=\int y \widetilde q(y\mid a,s)g^*(a\mid s)q_0(s)\,\mathrm{d}\mu_Y(y)\mathrm{d}\mu_\Acal(a)\mathrm{d}\mu_\Scal(s)\end{align} 
for some given target $g^*(a\mid s)$ (possibly signed). This encapsulates both treatment effect and policy values.

The stopping time may be prefixed, $\tau=T$ for a given constant $T$. Alternatively, we may stop adaptively, for example by following a fixed-confidence best-arm-identification algorithm such as Track-and-Stop \citep{garivier2016optimal}.
\end{example}

As a generalization of the inference task, one might consider inference for an adaptive target parameter in which case the mapping $\theta$ is itself a $\mathcal{F}_\tau$-measurable random functional. One such prominent example is inference for the value of the estimated best arm under a multi-armed bandit design. In this paper, we primarily consider marginal inference, that is the construction of confidence sets that contain the truth marginally over the randomness of the observations vectors, the design, the stopping time and the target parameter. Another setting of interest is conditional inference, where coverage is required to hold conditional on some or all elements of the triple (design, stopping time, target).

\subsubsection{What kind of inference?}

In the marginal inference setting, we want a confidence set $C_\tau$ that is $\Fcal_\tau$-measurable such that $\PP(\theta(q)\in C_\tau)\geq1-\alpha$ and $\PP(\theta(q)\notin C_\tau)$ is small and vanishing for $\theta'\neq\theta(q)$. Often this is already a byproduct of many bandit algorithms and/or their analysis, since these usually rely on always-valid confidence intervals on arm parameters, either explicitly or implicitly. Consider the upper confidence bound (UCB) algorithm for multi-arm bandits. Its analysis relies on proving that the confidence bound is in fact valid at all times with probablity almost 1 (see, \eg, the analysis of the ```good' event" in the proof of theorem 7.1 in \citealp{lattimore2020bandit}). As a consequence, we could simply use this confidence set (possibly with symmetric analysis for a lower bound), even at any adaptive stopping time. However, these sets are generally very big, as they are based on conservative concentration inequalities and union bounds, where the aim is often to get the right rates in problem parameters. Therefore, they tend to over-cover and under-reject. 

An approach that may be more appealing is, given some estimate $\hat\theta_\tau$, characterize the distribution of $\hat\theta_\tau-\theta(q)$ in some appropriately scaled asymptotic limit and then invert it. In the non-adaptive case, this is generally done via CLT, which yields Wald confidence intervals with miscoverage approaching exactly $\alpha$ and squared size shrinking rapidly as the limiting scaled variance over sample size. A hope is that we can use MCLT to obtain similar asymptotic normality also in the adaptive case and construct similar Wald confidence intervals. In \cref{sec:why_normality_fails} we discuss when MCLT breaks or does not break for standard estimators. In \cref{sec: reweighting} we discuss when and how we can modify standard estimators using weights in order to fix MCLT. In \cref{sec: always valid} we revisit the above-mentioned always-valid confidence intervals and discuss at least calibrating their anytime coverage to $1-\alpha$. In \cref{sec: non-normal} we discuss characterizing and inverting limiting distributions even when not normal.

\section{Why does adaptivity ruin asymptotic normality? (And when it doesn't)}\label{sec:why_normality_fails}

The standard approach to characterizing the asymptotic distribution of an estimator at a fixed time $T$ under adaptively collected data is to write it as a martingale difference sequence (MDS) and to apply a martingale CLT (MCLT).
While there are many MCLTs, they almost all essentially rely on a similar pair of sufficient conditions: (1) asymptotic negligibility of individual terms and (2) quadratic variation convergence (or conditional variance convergence, which is equivalent under mild conditions, see theorem 2.23 in \citealp{hall2014martingale}). For concreteness, we state below a standard MCLT, a slightly simplified rewriting of theorem 3.2 in \cite{hall2014martingale}. Their result applies to an MDS array $\{ X_{t,T} : t\in [T] \}$ where $(X_{t,T})$ is adapted to the filtration $(\mathcal{F}_t)_{t \geq 1 }$. The result requires the following two conditions.

\begin{assumption}[Asymptotic negligibility]\label{asm:asymp_negl}
    $\max_{t\in [T]} |X_{t,T}| \xrightarrow{p} 0$ as $T \to \infty$ and $E[\max_{t \in [T]} X_{t,T}^2]$ is bounded in $T$. 
\end{assumption}

\begin{assumption}[Quadratic variation convergence]\label{asm:qvar_conv}
    $\sum_{t=1}^T X_{t,T}^2 \xrightarrow{p} 1$.
\end{assumption}
\begin{theorem}[Simplified version of theorem 3.2 in \citealp{hall2014martingale}]
    Suppose that assumptions \ref{asm:asymp_negl} and \ref{asm:qvar_conv} hold. Then $\sum_{t=1}^T X_{t,T} \xrightarrow{d} \mathcal{N}(0,1)$.
\end{theorem}

\subsection{Quadratic variation convergence}

In this section we try to shed some light on the role of the quadratic variation convergence condition for asymptotic normality and how it may fail. The condition essentially requires conditional variances to be predictable. Intuitively, adaptivity can break this condition because we cannot predict how our design will adapt to future observations.

\subsubsection{Failure of the condition} It is easy to construct an example in which asymptotic normality fails by picking an estimator that can be written as the mean of an MDS and by choosing the design so that the sum of conditional variances is unpredictable and quadratic variation does not converge.

Consider an explore-then-commit 2-armed bandit design, in which the experimenter randomizes equiprobably treatment assignment in the first $t_0$ rounds and then, in rounds $t=t_0+1,\ldots,T$ assigns the arm with highest empirical mean computed from the first $t_0$ rounds with probability $1-\epsilon$ and the other arm with probability $\epsilon$, for some $\epsilon \in (0,0.5)$. Suppose both arms have variance $1$ and same mean $\mu$. In our notation, $g_t(a_t \mid a_t^-) = 0.5$ for every $t=1,\ldots,t_0$ and $g_t(a_t \mid a_t^-) = \epsilon +  (1-2\epsilon) \bm{1}\{ a = \arg\max_{a'=1,2} \widehat{\mu}_{a',t_0} \}$, where $\widehat{\mu}_{a',t_0}$ is the empirical mean of arm $a'$ computed from the first $t_0$ rounds. Consider the inverse probability weighting (IPW) estimator of the mean of arm 1 defined as 
\begin{align*}
\widehat{\mu}_{1,T}^{\mathrm{IPW}} = \frac{1}{T} \sum_{t=1}^T \frac{\bm{1}\{A_t=1\}}{g_t(1 \mid A_t = 1)} L_t
\end{align*}
The fact that both arms have the same mean makes it equally likely in rounds $t_0+1,\ldots,T$, that the probability that choosing a given arm is $1-\epsilon$ or $\epsilon$, and therefore that conditional variance of the $t$-th term is $(1-\epsilon)^{-1}$ or $\epsilon^{-1}$.
It is then straightforward to show that, with $t_0$ fixed and $T$ growing, the IPW estimator converges to a mixture of two normals, each corresponding to one of these two equiprobable cases:
\begin{align*}
\sqrt{T} (\widehat{\mu}_{1,T}^{\mathrm{IPW}} - \mu) \xrightarrow{d} \frac{1}{2} \mathcal{N}\left(0, 2 +  \frac{(1-\epsilon)^2}{2} \right) + \frac{1}{2} \mathcal{N}\left(0, 2 +  \frac{\epsilon^2}{2} \right)
\end{align*}

\subsubsection{The stabilizing effect of dividing by arm sample size} One might notice that in the above example the sum of of conditional variances scales as the sample size of the arm. This suggests that to obtain a stable asymptotic variance under adaptivity we can divide the sum of outcomes by the realized arm sample size $N_1(T)=\sum_{t\leq T}\bm{1}\{A_t=1\}$, which forms the arm empirical mean (EM):
$$
\widehat{\mu}_{1,T}^{\mathrm{EM}}=\frac1{N_1(T)}\sum_{t=1}^T\bm{1}\{A_t=1\} L_t.
$$
It is indeed straightforward to check that in the above example, the empirical mean not only has deterministic asymptotic variance, but is also asymptotically normal:
\begin{align*}
\sqrt{N_1(T)} (\widehat{\mu}_{1,T}^{\mathrm{EM}} - \mu) \xrightarrow{d} \mathcal{N}(0,1).
\end{align*}
Asymptotic normality of the empirical mean is even preserved for any sublinear $t_0 = T^\alpha$, $\alpha \in (0,1)$ duration of the exploration phase. 

\subsubsection{Too much adaptivity breaks asymptotic normality of the empirical mean} Even a vanishing amount of adaptivity (\ie, $t_0$ constant) was enough to break the asymptotic normality of IPW, while EM remained asymptotically normal. However, under a significant amount of adaptivity, such as when the exploration period equal to a constant fraction of the total sample size, the asymptotic normality of EM breaks as well.
If we set $t_0=T/2$, then one can show
there exists three independent standard normal random variables $Z_1$, $Z_2$ and $Z_3$ such that
 \begin{align*}
\sqrt{N_1(T)} (\widehat{\mu}_{1,T}^{\mathrm{EM}} - \mu_1) \xrightarrow{d} \frac{ \frac{1}{\sqrt{2}} Z_1 + \left(\bm{1}\{ Z_1 > Z_2\} \sqrt{1-\epsilon} + \bm{1}\{Z_1 \leq Z_2\} \sqrt{\epsilon}\right) Z_3 }{\bm{1}\{ Z_1 > Z_2\} \sqrt{\frac{1}{2} + 1 - \epsilon}  +  \bm{1}\{ Z_1 \leq Z_2\} \sqrt{\frac{1}{2} + \epsilon } },
\end{align*}
which can readily be checked to be non-normal.

\subsubsection{A tighter sufficient condition for asymptotic normality under adaptivity} The behavior of the empirical mean under different amounts of adaptivity raises the question of what is a good condition to determine when a self-normalized martingale will be a asymptotically normal. Theorem 3.4 in \cite{hall2014martingale} gives such a sufficient condition. We present their result and condition in slightly simplified setting. Let $\{X_{t,T} : t \in [T]: T \geq 1\}$ be a MDS array such that $(X_{t,T})_{t \in [T]}$ is adapted to a filtration $(\mathcal{F}_t)_{t \geq 1}$. Their condition is that there exists a sequence $(u_T)_{t \geq 1}$ adapted to a filtration $(\mathcal{G}_T)_{T \geq 1}$ such that (i) the quadratic variation $U_T^2 = \sum_{t=1}^T X_{t,T}^2$ converges in probability to $u_T$:
\begin{align*}
    U_T^2 - u_T^2 = o_P(1),
\end{align*}
and that (ii) the martingale property is mostly preserved under the enriched filtration $(\mathcal{F}_t \vee \mathcal{G}_T)_{t \in [T]}$:
\begin{align*}
    \sum_{t=1}^T E[X_{t,T}\mid \mathcal{F}_{t-1}, \mathcal{G}_T ] = o_P(1) \qquad \text{and} \qquad \sum_{t=1}^T E[X_{t,T}\mid \mathcal{F}_{t-1}, \mathcal{G}_T ]^2 = o_P(1).
\end{align*}
Under an additional typical asymptotic negligibility condition and a condition that says that $u_T$ is asymptotically non-degenerate, the self-normalized martingale $U_T^{-1} \sum_{t=1}^T X_{t,T} \xrightarrow{d} \mathcal{N}(0,1)$. In the context of adaptive experiments, the condition essentially requires that the total quadratic variation doesn't contain too much information on the sequence of realized outcomes. We believe it is a relatively tight condition as it accounts for the asymptotic behavior of the empirical mean in the examples we've seen in this section: the condition holds in the sublinear exploration fraction case, but fails in the linear fraction case.

\subsection{Asymptotic negligiblity condition}

In the case of the sample mean, conditional variance of any individual observation is always negligible in front of the total quadratic variation. A situation where this is not trivially the case anymore is that of inverse propensity weighted (IPW) estimators, where conditional variance of any individual observation scales as the inverse probability of treatment arm allocation conditional on the history. We can indeed construct examples where asymptotic negligibility fails to hold by simply creating a situation where the exploration probability is of order $T^{-1}$ for some time point $t$. Yet, it seems to us that for designs for which IPW estimators are applicable, that is designs that enforce a non-zero exploration probability at every rounds such as Thompson sampling, exploration rate decay is usually slower, of the order or $T^{-1/2}$. We can check that for sublinear exploration rate decay, the asymptotic negligibility condition will always hold for terms of IPW estimators.

\section{Recovering normality at a fixed stopping time via observation weighting}\label{sec: reweighting}

One way that has been proposed to recover normality when interested in inference at a fixed stopping time $\tau=T$ is to introduce weights that stabilize the conditional variances and make them converge \citep{bibaut2021post,hadad2021confidence,zhan2021off}. Consider the contextual bandit setting (\cref{ex:cb}, which nests \cref{ex:mab,ex:batch}). 
For brevity, let $S_t^-=(S_1,A_1,Y_1,\dots,Y_{t-1})$ and let $g_t$ refer to $S_t^-$-measurable random conditional density $g_t(\cdot\mid S_t=\cdot,S_t^-)$.
Suppose we have a (recentered) influence function $\phi(s,a,y;\eta,g)$ such that for any $\eta\in\Hcal$ and any conditional density $g\in\Gcal$ in some valid realization set for our designs (\ie, $\PP(g_t\in\Gcal\;\forall t\leq T)=1$),
$$
\theta(q) = \int \phi(s,a,y;\eta,g) \widetilde q(y\mid s,a)g(a\mid s)q_0(s)\,\mathrm{d}\mu_Y(y)\mathrm{d}\mu_\Acal(a)\mathrm{d}\mu_\Scal(s).
$$
For example, for 
the parameter $\theta(q)$ as in \cref{eq:policy value}, we can use
\begin{align}\label{eq:cb dr if}
\phi(s,a,y;\eta,g)=\int\eta(\widetilde a,s)\widetilde g(\widetilde a\mid s)\,\mathrm{d}\mu_\Acal(\widetilde a)+\frac{\widetilde g(a\mid s)}{g(a\mid s)}(y-\eta(a,s)).
\end{align}
With the appropriate choice of $\phi$ can also tackle, for example, quantiles, expectiles, superquantiles, \etc

Then, letting $\mathcal{F}_t = \sigma(S^-_t)$, for any pair of $(\mathcal{F}_t)_{t \geq 0}$-measurable sequences $\omega_t,\eta_t$, we can consider the estimator
$$
\hat\theta = \frac{\Gamma_T}T\sum_{t=1}^T\omega_{t-1}\phi(S_t,A_t,Y_t;\eta_{t-1},g_t),\quad\Gamma_T=\prns{\frac1T\sum_{t=1}^T\omega_{t-1}}^{-1}.
$$
By construction, $X_t=T^{-1/2}\omega_{t-1}(\phi(S_t,A_t,Y_t;\eta_{t-1},g_t)-\theta(q))$ is an MDS with respect to $(\mathcal{F}_t)_{t \geq 0}$. So to establish the asymptotic normality of $\Gamma_T^{-1}\sqrt{T}(\hat\theta-\theta(q))=\sum_{t=1}^T X_t$, it suffices to establish \cref{asm:asymp_negl,asm:qvar_conv}.

One approach inspired by the work of \citet{luedtke2016statistical} is to choose \begin{equation}\label{eq:std weighting}\omega_{t-1}\approx \mathrm{Var}^{-1/2}(\phi(S_t,A_t,Y_t;\eta_{t-1},g_t)\mid S^-_t).\end{equation}
That is, make our estimator an average of unbiased terms, each with conditional variance approximately one.

\citet{bibaut2021post} consider doing exactly this for the case of the parameter in \cref{eq:policy value} in a contextual bandit (\cref{ex:cb}) using the (recentered) influence function in \cref{eq:cb dr if}.
They show that \cref{asm:qvar_conv} is satisfied as long as the error in the ``$\approx$" above converges to 0 almost surely and that \cref{asm:asymp_negl} is satisfied as long as $g_t$ is lower bounded by $\Omega(t^{-1/2})$ almost surely. They go on to show how exactly to come up with conditional variance estimators to achieve \cref{eq:std weighting}. This appears nontrivial because we observe only one draw of the term whose variance we want to estimate. This is resolved by noting that the only thing changing between observations is our design $g_t$, so one can employ importance sampling (that is, on top of the importance sampling in $\phi$ itself). Using these conditional variance estimates, they obtain asymptotic normality. 

Regarding $\eta_t$, any convergent sequence suffices for asymptotic normality, but one that approximates $\int y\widetilde q(y\mid s,a)\,\mathrm{d}\mu_Y(y)$ can achieve lower asymptotic variance. \citet{bibaut2021risk} provide sharper guarantees on fitting this regression with importance-sampling-weighted nonparametric least squares in cases where the importance weights diverge.

\citet{hadad2021confidence} consider the special case of $K$-armed bandits (\cref{ex:mab}) with $\theta(q)=\int y\widetilde q(y\mid k)\,\mathrm{d}\mu_Y(y)$ being the mean of some arm $k$ and $\phi(a,y;\eta,g)=\eta+\frac{\bm1\{a=k\}}{g(k)}(y-\eta)$. They show that $\omega_{t-1}=\sqrt{g_{t}(k)}$ suffices to establish the conditions of MCLT. That is, no explicit estimation of conditional variances is required. They go on to consider a class of weighting strategies that make variances predictable and MCLT go through, including square-root-propensity weighting, and they propose a heuristic that aims to minimize asymptotic variance in this class.

In the contextual bandit OPE setting, \cite{zhan2021off} propose an alternative weighting scheme to that of \cite{bibaut2021post} that doesn't require outcome models estimates in the weights. Specifically, they use weights of the form
\begin{align}
    \omega_t = \sum_{a=1}^K \frac{\widetilde{g}^2(a \mid S_t)}{g_t(a \mid S_t, S_t^-)}.
\end{align}
Their guarantees rely on an alternative set of assumptions than that of \cite{bibaut2021post}. One key difference is that they require that 
\begin{align}
\sup_{s,a} \left\lvert \frac{g_t^{-1}(a \mid s, S_t^-)}{E[g_t^{-1}(a \mid s, S_t^-)]} - 1\right\rvert \to 0~~ a.s,
\end{align}
which at a high-level says that the expected design at should stabilize to its expectation over many runs of the experiment. This will usually hold under most best-arm identification or regret-minimizing contextual bandit algorithms if for every $s$, the conditional arm means are sufficiently separated from each other. This, however, may fail to hold in the so-called ``exceptional regime'' of optimal dynamic treatment estimation \citep{luedtke2016statistical}, in which there exists a non-zero-probability strata of the context space where two arms have the same conditional mean. In such a regime, common bandit algorithms tend to play either of these two arms at random on such a strata, jeopardizing the design stabilization condition. While \cite{bibaut2021post} do not require such a condition, they pay a price for using their importance-sampling device, by requiring entropy conditions on the set $\mathcal{G}$ of possible realizations of the design.

Another application of inverse estimated conditional standard deviation weighting is the procedure for M-estimation of parametric model under adaptive data collection by \cite{zhang2021statistical}. They show that by maximizing an inverse-conditional-standard-deviation-weighted empirical risk, you can obtain an asymptotically normal estimate of the finite dimensional parameter of interest, provided it is well specified. In the misspecified regime, we may instead wish to simply minimize a criterion (such as squared error or policy loss) over a hypothesis class, where unattenuated importance-sampling weights should be used to ensure we target the right average criterion \citep{bibaut2021risk}.

\section{Confidence intervals via always-valid inference}\label{sec: always valid}

In this and the next section we turn to approaches that sidestep asymptotic normality altogether to enable inference. First we discuss the role of always-valid inference.

\subsection{Multi-armed bandits}

The solution of reweighting the IPW estimator for mean arm reward in the case of multi-armed bandits (\cref{ex:mab}) works for a fixed stopping time, $\tau=T$. However, in the case of multi-armed bandits (\cref{ex:mab}), the issue of adaptive treatment allocation and adaptive stopping are essentially the same. So, solutions that can deal with adaptive stopping, such as always-valid inference, offer an alternative fix without relying on asymptotic normality. 

Consider the case of \cref{ex:mab} with any parameter $\theta(q)$ that only depends on the density $\widetilde q(y\mid k)$ of the $k$-th arm, such as the mean of arm $k$, $\theta(q)=\theta^{(k)}(q)=\int y\widetilde q(y\mid k)\,\mathrm{d}\mu_Y(y)$. 
Since we are only interested in arm $k$, both the design and the densities of other arms are ancillary. That is, we should only care about observations from pulling arm $k$. Let $t_k(s)=\inf\{t:
N_t(k)\geq s\}$ be the time of the $s$-th draw from arm $k$. Then, the sequence $Y_{t_k(1)},Y_{t_k(2)},\dots$ actually consists of i.i.d. draws from arm $k$. The only problem now is that at inference time the number of draws available from this sequence, $N_\tau(k)$, is adaptive, capturing both the potential adaptivity of treatment allocation and/or of stopping in one.

The literature on always-valid inference studies constructing confidence intervals that are simultaneously valid at all times -- equivalently, at adversarial adaptive stopping times. We say that a sequence $(\mathcal{C}_t)_{t \geq 1}$ is a joint confidence sequence for a parameter $\theta$ if $\PP(\exists t \geq 1 : \theta \not\in \mathcal{C}_t ) \leq \alpha$. Tools for constructing joint confidence sequences for $\theta_k(q)$ can be directly applied to the above problem to handle the adaptivity of the effective stopping time $N_\tau(k)$. In particular, if $\PP(\exists t \geq 1 : \theta^{(k)}(q) \not\in \mathcal{C}^{(k)}_t ) \leq \alpha$ then $\PP(\exists t \geq 1 : \theta^{(k)}(q) \not\in \mathcal{C}^{(k)}_{N_\tau(k)} ) \leq \alpha$, where $\mathcal{C}^{(k)}_{N_\tau(k)}$ is the interval we have at time $\tau$ based on all data so far. For the difference between two arms $k,k'$ (the average treatment effect), we can note that $\PP(\exists t \geq 1 : \theta^{(k)}(q)-\theta^{(k')}(q) \not\in \mathcal{C}^{(k,k')}_\tau ) \leq 2\alpha$, where $\mathcal{C}^{(k,k')}_\tau=\{\theta^{(k)}_0-\theta^{(k;)}_0:\theta^{(k)}_0\in\mathcal{C}^{(k)}_{N_\tau(k)},\theta^{(k')}_0\in\mathcal{C}^{(k)}_{N_\tau(k')}\}$.

Construction of exact confidence sequences for a sequence of estimators $(\widehat{\theta}_t)_{t \geq 1}$ that can be written as means of a MDS (that is, $\widehat{\theta}_t-\theta = t^{-1} \sum_{s=1}^t X_s$ where $X_1,X_2,\dots$ form an MDS) can be achieved under knowledge of bounds on the conditional moment generating function. Such bounds can be obtained by direct application of union bounds with a given $\alpha$-spending schedule. More refined, state-of-the-art always-valid sequences with can be found in \cite{howard2021time}, which use a novel stitching technique.

While exact always-valid confidence sequences are theoretically appealing, they tend to be very wide in practice, and they do require knowledge of bounds on the MGF, which are usually not available, or when they are, tend to be very conservative. One of the reasons we aspire to asymptotic normality and sought to recover it in the above is that it allows for exact calibration of coverage probability (asymptotically).

An alternative to exact always-valid inference is to leverage an invariance principle. This is a sequential notion of asymptotic normality: that under appropriate conditions $\sum_{s=1}^t X_s$ is approximated by Brownian motion up to $o(\sqrt{t})$ error. \citet{bibaut2022near,waudby2021time} employ such tools to approximate the problem by that of testing the drift of a Brownian motion, which can be done exactly. In particular, \citet{bibaut2022near} show that requiring coverage only for $t\geq t_0$ (equivalently, setting $\Ccal_t=\RR$ for $t<t_0$) and treating the problem as testing Brownian motion drift thereafter is actually optimal in terms of the expected time to exclude a false value from the confidence sequence. Optimality, however, is when we require coverage against an adversarial adaptive stopping time. Since here we are dealing with a particular adaptive stopping time, coverage may still be conservative. That is, unless the particular adaptive stopping time coincides with the worst case. For example, if stopping is based exactly on when the very same confidence sequence excludes $0$, then it is also the worst case for it incorrectly excluding $0$. Therefore, if always-valid coverage is roughly $1-\alpha$, as would be the case with the procedure of \citet{bibaut2022near}, then there is no conservativeness. That is, stopping based on a well-tuned always-valid procedure for testing a given hypothesis does not need any reanalysis by a secondary inferential procedure if it is for the same hypothesis.

\subsection{Contextual bandits}\label{sec: cb anytime}

The reweighting solution in \citet{bibaut2021post} works for a fixed stopping time, $\tau=T$. However, that same solution can be combined with always-valid inference to handle any adaptive stopping time. In particular, \citet{bibaut2021post} proceed by showing that their stabilization allows the estimator to satisfy \cref{asm:asymp_negl,asm:qvar_conv}. The assumptions needed to invoke the invariance principle in \citet{bibaut2022near} are similar (albeit slightly stronger). Essentially, under similar (but slightly stronger) conditions to theorems 1 and 2 of \citet{bibaut2021post}, the sequence of observations of $X_t=\omega_{t-1}(\phi(S_t,A_t,Y_t;\eta_{t-1},g_t)-\theta(q))$ satisfies the conditions of theorems 1 and 4 of \citet{bibaut2022near} for testing that $X_t$ is an MDS. The strategy is similar to section 6 of \citet{bibaut2022near}. Therefore, combining the two approaches we are able to handle always-valid inference on policy value using data sequentially collected by a running contextual bandit. As before, however, if we are interested in a particular adaptive stopping time rather than an adversarial one (equivalently, all stopping times simultaneously) then we may be too conservative.

\section{Embracing non-normality: inverting a non-trivial distribution}\label{sec: non-normal}

Inference methods we have reviewed so far try to eschew dealing with the potentially complex distribution of test statistics under adaptive sampling. In order to so, they either aim to recover normality at the target stopping time or to ensure validity at every stopping time such as by relying on joint normality of an asymptotic limiting sequence. It should come as no surprise that trying to ensure validity at every stopping time is inefficient if one cares about inference at a given stopping time. We also argue that forcing normality via weighting loses some information that could be leveraged to achieve greater power. To overcome these limitations requires characterizing potentially non-trivial distribution of test statistics.

\subsection{Bernoulli simulation: inverting an arbitrary test}\label{sec:bernoulli_sim}

Let us start by giving an example that illustrates that, in theory, we don't need asymptotic normality to conduct inference at a given stopping time. Suppose we are in the $K$-armed bandit setting (\cref{ex:mab}) where outcomes are Bernoulli with means $\bm{\mu} = (\mu_1,\ldots, \mu_K)$. Let $q_{\bm\mu}(y\mid a)=\mu_a^y(1-\mu_a)^{1-y}$ be the factor $q$ of the data-generating distribution as defined in our set-up corresponding to Bernoulli arms with mean vector $\mu$. Suppose we run an arbitrary adaptive design $\bm g = (g_1, g_2,\ldots)$ and we stop at stopping time $\tau$. We care about confidence intervals for a parameter $\theta(q)$, or equivalently $\widetilde \theta(\bm\mu)$. Let $\rho$ be an $\bar O_\tau$-measurable test statistic. Then, following \citet{wei1990statistical}, a procedure to generate a $1-\alpha$ confidence set $\mathcal{C}$ is as follows:\footnote{We thank P. Aronow for pointing us to this procedure.}
\begin{itemize}
    \item For every $\bm\mu\in[0,1]^K$ let $\kappa_{\bm\mu}(1-\alpha)$ be the $(1-\alpha)$-quantile of $\rho$ under
     the (known) data density $g_1(a_1)q_{\bm\mu}(y_1\mid a_1)g_2(a_2\mid a_2^-)q_{\bm\mu}(y_2\mid a_2)\cdots$. 
    \item Let $\mathcal{C} = \left\lbrace \bm\mu : \rho \leq \hat \kappa_{\bm\mu}(1-\alpha) \right\rbrace$.
\end{itemize}
This computationally-infeasible procedure can be (still computationally intensively) approximated by taking a regular grid of $[0,1]^K$ and for each grid point $\bm\mu$ take the empirical quantile of many repeated draws of $\bar O_\tau$ generated by simulating the bandit system with the given algorithm ($\bm g,\tau$) using $q_{\bm\mu}$ to generate arm rewards at each round.

This procedure ensures that $\mathcal{C}$ is a $1-\alpha$ confidence set for the vector $\bm\mu$. Marginalizing it to $\widetilde \theta(\mathcal{C}) = \left\lbrace  \widetilde \theta(\bm\mu) : \bm\mu \in \mathcal{C} \right\rbrace$ yields a $1-\alpha$ confidence interval for $\theta(q) = \widetilde \theta(\bm\mu)$. In particular, \citet{wei1990statistical} specifically focus on inference after a play-the-winner algorithm and consider a marginalized interval for $\widetilde\theta(\bm\mu)=\mu_2-\mu_1$ in a two-arm case using $\rho=\hat\mu^{\mathrm{EM}}_{2,T}-\hat\mu^{\mathrm{EM}}_{1,T}$, where $\widetilde \theta(\mathcal{C})$ can be written as $\{\Delta\in[-1,1]:\rho\leq\sup_{\mu_2\in[0\vee\Delta,1+0\wedge\Delta]}\kappa_{\mu_2-\Delta,\mu_2}(1-\alpha)\}$.

Whether $\widetilde \theta(\mathcal{C})$ is optimally tight depends on the specific test statistic we invert. While characterizing optimality of tests under general data-generating distributions and adaptive designs is tough, for tests of simple-vs-simple hypotheses the Neyman-Pearson lemma characterizes optimal tests, and for one-sided hypothesis, the theory of uniformly most powerful tests is well established under exponential families (see e.g., \citealp{keener2010theoretical}). We could leverage these theories if we could show some form of asymptotic equivalence of test statistic distributions with a tractable distribution. We review some works that do precisely this in subsection \ref{sec:asymp_theory_xps}. 

In the same spirit as the Bernoulli simulation we presented above, in their pioneering work, \cite{rosenberger1999bootstrap} tackle the same Bernoulli setting with a slightly different method. They first obtain arm mean estimates $\widehat{\bm{\mu}} = (\widehat \mu_1, \ldots, \widehat \mu_K)$. They compute quantiles of the components of the empirical arm means by running the experiment many times under density $g_1(a_1) q_{\widehat{\bm{\mu}}}(y_1 \mid a_1) g_2(a_2 \mid a_2^-) q_{\widehat{\bm{\mu}}}(y_2 \mid a_2) \ldots $, and use these quantiles to form confidence intervals. 

\subsection{Asymptotic distribution of a test statistic in the asymptotic experiment}\label{sec:asymp_theory_xps}

\cite{hirano2023asymptotic} and \cite{adusumilli2023optimal} develop an analog of the limit theory of experiments \citep{van2000asymptotic} for sequential experiments.

While \cite{hirano2023asymptotic}'s results specialize to the batch bandit setting, \cite{adusumilli2023optimal} provides slightly different results that seem to apply to more general settings. We start by summarizing \cite{hirano2023asymptotic}'s contribution in the batch bandit setting in our notation. 

\subsubsection{Optimal tests in a limit batch bandit experiment}

Consider the batch multi-armed bandit setting (\cref{ex:batch}), with at most $b_{\max}$ batches, where each batch has size $n$. While \cite{hirano2023asymptotic}'s results should readily extend to adaptive selection of the number of batches, they present their results for a fixed number of batches $b_{\max}$. Following our notation, let $q$ be a conditional density of outcomes given actions, let $\widetilde{\bm{g}} = (\widetilde g_1, \ldots, \widetilde g_{b_{\max}})$ be the sequence of adaptive batch designs, and let $P^{(n)}_{q, \widetilde{\bm{g}}}$ be the corresponding distribution of the data sequence $\bar O_{b_{\max}}^{(n)}  = (O_1^{(n)},\ldots,O_{b_{\max}}^{(n)})$. 

\paragraph{The limit theory of experiments setting} Following the setting of the limit theory of experiments, the authors consider a sequence of experiments indexed by the batch size $n$. While their results extend to the semiparametric setting, they are easiest to present under a parametric model for $q$. Specifically, suppose that $q$ belongs to a parametric model $\{ q_\theta : \theta \in \Theta\}$ where $\Theta \subset \mathbb{R}^d$. The limit theory of experiments considers a null hypothesis $H_0 : q = q_\theta$ and a sequence of ``local alternatives'' indexed by $n$ of the form $H_{1,n}(h) :  q = q_{\theta + h / \sqrt{n}}$ for some $h \in \mathbb{R}^d$. Alternatively, we can say that we consider the sequence of hypotheses $(H_{0,n}, H_{1,n}(h))$ with $H_{0,n} : \bar O^{(n)}_{b_{\max}} \sim P^{(n)}_{q_{\theta}, \widetilde{\bm{g}}}$ and $H_{1,n}(h) : \bar O^{(n)}_{b_{\max}} \sim P^{(n)}_{q_{\theta + h / \sqrt{n}}, \widetilde{\bm{g}}}$. Let $\Pi_b^{(n)} = (\Pi_{b, a}^{(n)} : a \in [K])$ where $\Pi_{b, a}^{(n)}$ is the fraction of draws of arm $a$ in batch $b$ and let $S^{(n)}_b$ be a statistic computed from the observed data up to batch $b$.

\paragraph{Main result: characterization of the limit experiment} Similarly to classical results in the i.i.d. setting, the authors show that under quadratic mean differentiability (QMD) of $\theta \mapsto q_\theta(\cdot \mid a)$ at for every $a$, the distribution of the test statistics under $q_{\theta + h / \sqrt{n}}$ converges to the distribution of some ``randomized'' test statistics under a limit normal data experiment. Specifically, they show that there exists fixed functions $(\widetilde{\Pi}_{b,a}, \widetilde{S}_b: a \in [K], b \in [b_{\max}])$, and random variables $(U, Z_{b,a} : a \in [K], b \in b_{\max}])$ such that
\begin{enumerate}
    \item $U \sim \mathrm{Uniform}([0,1])$,
    \item For every $b=1,\ldots,b_{\max}$, conditional on $(U, Z_1,\ldots, Z_{b-1})$, $Z_{b,1},\ldots, Z_{b,K}$ are independent and
    \begin{align*}
        Z_{b,a} \mid U, Z_{1}, \ldots, Z_{b-1} \sim \mathcal{N}(\Pi_{b,a}  I_a(\theta) h, \Pi_{b,a}  I_a(\theta)),
    \end{align*}
    where $\Pi_{b,a}$ is the fraction of units assigned to arm $a$ in batch $b$ of the limit experiment, defined as 
    \begin{align*}
        \Pi_{b,a} = \widetilde \Pi_{b,a}(U,Z_1,\ldots,Z_{b-1}),
    \end{align*}
    and $I_a(\theta)$ is the $d \times d$ Fisher information matrix at $\theta$ of the model $\{q_\theta(\cdot \mid a) : \theta \in \Theta\}$.
    \item Under the sequence of local alternatives $H_{1,n}(h)$, the statistics
    \begin{align*}
        (\Pi_{1}^{(n)}, S_1^{(n)}, \ldots, \Pi_{b_{\max}}^{(n)}, S_{b_{\max}}^{(n)}) \rightsquigarrow^{h} (\Pi_1, S_1, \ldots, \Pi_{b_{\max}}, S_{b_{\max}}),
    \end{align*}
    where the limit arm fractions $\Pi_1,\ldots \Pi_b$ are defined above, and the outcome statistics $S_1,\ldots, S_b$ are defined by $S_{b} = \widetilde S_{b}(U,Z_1, \ldots Z_b)$.
\end{enumerate}
The collection of arm fractions $\Pi_1,\ldots, \Pi_B$ and conditionally normal random vectors $Z_1,\ldots,Z_b$ can be seen as defining a limiting batch bandit experiment. In particular, the asymptotic representation theorem we just presented says that the sequence of experiments with sequence of outcome models $(q_{\theta+h / \sqrt{n}})_{n \geq 1}$ converges to a limit experiment with conditional outcome models that are normal and are fully determined by $h$, the Fisher information matrices $I_a(\theta)$ and the limit arm fractions $\Pi_b$ as a function of previous batches. The vectors $\theta$ and $h$ are a user-chosen parameter that specifies the sequence of local alternatives $(H_{0,n},H_{1,n}(h))$ and the Fisher information matrix is also known for the given model and value of $\theta$ under consideration. The limit allocation rules $\Pi_b$ might be more complex to identify as the ART only guarantee their existence. \cite{hirano2023asymptotic} show that explicit convergence of the finite sample allocation rules $\Pi_b^{(n)}$ can for instance be established in the case of batched Thompson sampling.

\paragraph{Application to asymptotic power envelope calculation}

One benefit of the representation of the limit experiment is that it allows for characterization of optimal tests. In particular, the Neyman-Pearson lemma tells us that, in the limit experiment, the optimal test of the hypothesis $H_0 : h= 0$ against the alternative $H_1 : h = h_1$ for some value of $h_1$ is the likelihood ratio test (LRT). A simple example is when $d = K$ and each component of $\theta$ parametrizes a different arm (formally,  $q(\cdot \mid a)  = q^{(a)}_{\theta_a}(\cdot)$ for every arm $a \in [K]$), in which case $Z_{b,a}$ is one-dimensional and $Z_{b,a} \mid U, Z_1,\ldots, Z_{b-1} \sim \mathcal{N}(\Pi_{b,a} I_a(\theta_a) h_a, \Pi_{b,a} I_a(\theta_a))$. It is then straightforward to show that the corresponding likelihood ratio takes the closed form
\begin{align*}
    \frac{f_h(U,Z_1,\ldots, Z_{b_{\max})}}{f_0(U,Z_1,\ldots, Z_{b_{\max}})} = \prod_{a=1}^K \exp \left( h_a \sum_{b=1}^{b_{\max}} X_{b,a}  - \frac{1}{2} h_a^2  I_a(\theta_a) \sum_{b=1}^{b_{\max}} \Pi_{b,a} \right),
\end{align*}
which allows for simple implementation of the LRT in the limit experiment.
Simulating the LRT in the limit experiment allows for power envelope calculations. 

\paragraph{Asymptotic power comparisons} \cite{hirano2023asymptotic}
compare the asymptotic power of tests in a sequence of two-batch batched Thompson sampling experiments with two arms. They consider a sequence of experiments with outcome distribution parametrized by $\theta_n = \beta [1, 1]^\top + [h_1, h_2]^\top / \sqrt{n}$, and a sequence of tests of the hypothesis $H_0 : h_1 = h_2$.
The limit Thompson experiment is fully characterized by arms fractions in the first batch, the distribution of arm means $Z_1=(Z_{1,1},Z_{1,2})$ in the first batch, the arm fractions vector $\Pi_2=(\Pi_{2,1}, \Pi_{2,2})$ in the second batch, and the conditional distribution of arm outcomes $Z_2=(Z_{2,1}, Z_{2,2})$, which are given by
\begin{align*}
\Pi_{1,1} =& \frac{1}{2}, ~~ \Pi_{1,2} = \frac{1}{2}, ~~
    \Pi_{2,2}(Z_1) = \Phi\left( \frac{Z_{1,2} -  Z_{1,1}}{\sqrt{2 (\sigma_0^2 + \sigma_1^2)}} \right) ,~~ \Pi_{2,1}(Z_1) = 1 - \Pi_{2,2}(Z_1),\\
    Z_1 \sim&  \mathcal{N} \left( 
    \begin{bmatrix} 
        h_1 \\ 
        h_2 
    \end{bmatrix}, 
    \begin{bmatrix}
        \sigma_1^2 & 0 \\
        0 & \sigma_2^2
    \end{bmatrix}
    \right),~~
    Z_2 \mid  Z_1 \sim  \mathcal{N} \left( 
    \begin{bmatrix}
        h_1 \\
        h_2
    \end{bmatrix},
    \begin{bmatrix}
        \frac{\sigma_1^2}{\Pi_{2,1}(Z_1)} & 0 \\
        0 & \frac{\sigma_2^2}{\Pi_{2,2}(Z_1)}
    \end{bmatrix}
    \right),
\end{align*}
where $\sigma_1^2$ and $\sigma_2^2$ are the inverse Fisher information of the arm outcome models. The difference-in-means statistic in the limit experiment is given by 
\begin{align*}
    \hat{T}_{\mathrm{DIM}} = \frac{\frac{1}{2} Z_{1,1} + \Pi_{2,1} Z_{1,2}}{\frac{1}{2} +  \Pi_{2,1}} - \frac{\frac{1}{2} Z_{1,2} + \Pi_{2,2} Z_{2,2}}{\frac{1}{2} +  \Pi_{2,2}}.
\end{align*}
As we have seen in section \ref{sec:why_normality_fails}    , due to the non-negligible fraction of outcomes that determine the sample size of each arm, this statistic is non-normal. Nevertheless, one can simulate its distribution under the null and calibrate the rejection threshold so as to control test size.
Another test statistic is the one given in \cite{zhang2000inference}, in which observations are conditional-variance-weighted so as to ensure normality:
\begin{align*}
    \hat{T}_{\mathrm{ZJM}} = \frac{Z_{1,2} - Z_{1,1}}{\sqrt{\sigma_1^2 + \sigma_2^2}} + \frac{Z_{1,2} - Z_{1,1}}{\sqrt{\frac{\sigma_1^2}{\Pi_{2,1}} + \frac{\sigma_2^2}{\Pi_{2,2}}}}.
\end{align*}
Simulation seems to indicate that the difference-in-means test uniformly outpeforms the conditional-variance-weighted test, suggesting that ensuring normality comes at the cost of loss of information. Furthermore, for every alternative, the difference-in-means test almost matches the power envelope (that is the power of the corresponding likelihood ratio test). 

\paragraph{\cite{adusumilli2023optimal}} The author similarly shows that under QMD of the environment response likelihood factor $q$ (along parametric models or parametric submodels in a nonparametric model), the limit experiment of a sequence of local experiments with $K$ treatment arms can be represented by $K$ Wiener processes each stopped at limit stopping times representing the asymptotic sample size fraction of each arm. At a technical level, similarly to \cite{hirano2023asymptotic}, the results rely on a sequential extension of Le Cam's third lemma. The specific proof technique used by \cite{adusumilli2023optimal} crucially leverages Girsanov's theorem to carry out the change of measure from the limit null to the limit alternative in the limit experiment. 

At a high-level, \cite{adusumilli2023optimal} explicitly formulates what we think is the key take-away of the limit analysis of local sequences of experiments, which \cite{hirano2023asymptotic} also seem to hint at. That is, in general, asymptotically optimal tests need only rely on the arm means of the scores or the efficient influence function of the parameters of interest, evaluated at the final arm sample sizes. While we are not aware of guarantees for this approach outside of the diffusion asymptotics regime (the asymptotic regime of a sequence of experiments with effect size scaling as $n^{-1/2}$), we formulate a tentative procedure for inference after adaptive data collection.

\subsection{A tentative general procedure for post-contextual bandit inference with adaptive stopping}

The solution for post-contextual bandit inference discussed in \cref{sec: reweighting} only works for fixed stopping time, $\tau=T$, and relies on a lower bound on the design at each time, conditional on the past. The solution discussed in \cref{sec: cb anytime} gives anytime-valid post-contextual bandit inference, which handles adaptive stopping times but can be conservative for a given adaptive stopping time. The results of \cite{adusumilli2023optimal} suggest the existence of limiting stopping time procedure that we need only simulate to obtain the (possibly non-normal) null distribution, which we can then invert. However, the result is not constructive and unless we know what this limiting stopping time is or how to approximate it, we cannot simulate this distribution. In the fully parametric case, the approach outlined in \cref{sec:bernoulli_sim} allowed us to exactly simulate the stopping time. Based on the results of \cite{adusumilli2023optimal} and the insights of \cref{sec:bernoulli_sim}, we conjecture that this simulation procedure can be leveraged in the semiparametric setting to yield approximately valid confidence intervals, at least in the case where $\tau$ diverges to $\infty$. We outline the tentative proposed procedure in the following.

\subsubsection{Statistical model and arm decomposition of the EIF}
We formulate it in the contextual bandit setting (\cref{ex:cb}) with $K$ arms, $\Acal=[K]$. Suppose that the repeated factors $q_0$ and $\widetilde q$ of the density are parameterized by a scalar target parameter $\theta \in \Theta \subset \mathbb{R}$ and a possibly infinite-dimensional nuisance parameter $\eta \in \Hcal$. That is, $q_0 = q_{0, \theta, \eta}$ and $\widetilde q = \widetilde q_{\theta, \eta}$ for $\theta \in \Theta$ and $\eta \in \Hcal$. 
Let $(s, a, y) \mapsto \psi(s, a, y; g, \theta', \eta)$ be the efficient influence function of $\theta$ 
at $(\theta', \eta)$ under fixed logging policy $g$.
Suppose that $\psi$ decomposes as 
\begin{align}
    \psi(s, a, y; g, \theta, \eta) = \sum_{a = 1}^K  \psi_a(s, y; g(a \mid \cdot), \theta, \eta).
\end{align}
Let $\widetilde \psi_a(s, y; \theta, \eta) = \psi_a(s, y; 1, \theta, \eta)$ be the function obtained from $\psi_a$ by setting $g(a \mid \cdot)$ to 1, that is, by setting $g$ to the logging policy that always assigns arm $a$. 

An example where these assumptions hold is off-policy evaluation in the two-arm contextual bandit (arms $a \in \{0,1\}$) model where $q_{0, \theta, \eta}$ does not depend on $\theta$ and where rewards are Bernoulli and their conditional probability mass function is parameterized as follows:
\begin{align}
    \widetilde q_{\theta, \eta}(1 \mid a, s) = \theta \bm{1}\{a=1\} + \widetilde q_{\eta}(1 \mid 0, s),
\end{align}
with $\theta \in (-1,1)$ and $\widetilde q_{\eta}(1 \mid 0, s) \in [0\vee(-\theta), 1 \wedge(1 - \theta)]$.

\subsubsection{Experiment and test statistic} Suppose we run the experiment with an adaptive design $\bm{g}=(g_1,g_2,\ldots)$ until stopping time $\tau$.
Let $\rho_a(\theta) = N_a^{-1} \sum_{t=1}^\tau  \bm{1}\{A_t = a\}\widetilde \psi_a(S_t, Y_t ; \theta, \widehat{\eta}_{t-1})$, where $\widehat{\eta}_{t-1}$ is a nuisance estimator based on observations from rounds $1,\ldots, t-1$, and where $N_a = \sum_{t=1}^\tau \bm{1}\{ A_t = a\}$ is the sample size of arm $a$ at the end of the experiment. Let our test statistic be $\rho(\theta) = \left\lvert \sum_{a=1}^K \rho_a(\theta) \right\rvert$

\subsubsection{Inference procedure}  For every $\theta \in \Theta, \eta \in \Hcal, \alpha \in (0,1)$, let $\kappa_{1-\alpha}(\theta, \eta)$ be the $(1-\alpha)$ quantile of $\rho(\theta)$ under density $q_{0,\theta, \eta}(s_1)g_1(a_1 \mid a_1^-) \widetilde q_{\theta, \eta}(y_1 \mid a_1, s_1) q_{0,\theta, \eta}(s_2)g_2(a_2 \mid a_2^-) \widetilde q_{\theta, \eta}(y_2 \mid a_2, s_2) \ldots$. A procedure we conjecture to yield approximately valid confidence intervals under mild assumptions is as follows.
\begin{enumerate}
    \item For every $\theta \in \Theta$, compute $\kappa_{1-\alpha}(\theta, \widehat{\eta}_{\tau})$. 
    \item Return the confidence set $\mathcal{C} = \{ \theta : \rho(\theta) \leq  \kappa_{1-\alpha}(\theta, \widehat{\eta}_{\tau})\}$
\end{enumerate}
Computing $\kappa_{1-\alpha}(\theta, \widehat{\eta}_{\tau})$ involves simulating many replications of the bandit experiment using the hypothesized arm distributions given by $\theta, \widehat{\eta}_{\tau}$ and running our algorithm given by $\bm g,\tau$ in this simulated environment.
The procedure sweeps over possible values of $\theta$ but uses the estimated nuisance parameter $\eta$. The error in estimation of $\eta$ should be relatively inconsequential if (1) $\tau$ is very large, (2) the EIF is robust to $\eta$ misspecification. The former will be the case if the effect size under the true DGP is very small.

\section{Concluding remarks}

Adaptive experiments are increasingly being used not just in settings where we desire a continually improving system but where a scientific question is of interest. Often we want to answer this question using confidence intervals with frequentist guarantees. Algorithms for adaptive experiments in fact usually leverage bounds with frequentist guarantees, but while useful for the design of algorithms and for proving regret rates, they can be far too conservative to be useful for data analysis in practice. The analogue in non-adaptive experiments would be to invert Hoeffding's inequality instead of CLT (only worse due to the many union bounds too).

Ideally, we would like to use asymptotic normality so we can obtain simple Wald-type confidence intervals with asymptotic coverage calibrated exactly to a desired level. However, adaptivity -- whenever it remains significant throughout the experiment rather than washing away -- breaks the asymptotic normality of standard estimators like sample mean and IPW. We reviewed when this happens and what we can do when it does, including reweighting to regain normality, always-valid inference, and inverting the non-normal distribution.

Many of these solutions are very effective, but this is certainly not a solved problem yet. Reweighting can give up some efficiency and it also may not deal with adaptive stopping times. Always-valid inference can be conservative because it is valid always rather than at the actual stopping time. It is not always possible to characterize limiting non-normal distributions, especially when departing batched settings. Therefore, lacking a silver bullet, care must be taken in practice in analyzing data from adaptive experiments. We hope this article offers the understanding needed to take such care as well as inspire further work to tackling this important problem.

\bibliography{refs}
\bibliographystyle{plainnat}

\end{document}